\documentclass[useAMS,usenatbib]{mn2e}
\usepackage{natbib}
\usepackage{threeparttable}
\usepackage{graphicx}% Include figure files
\usepackage{dcolumn}% Align table columns on decimal point
\usepackage{bm}% bold math
\usepackage{amsmath,ulem}
\usepackage{epsf}
\usepackage{color}
\usepackage{amssymb}
\usepackage{latexsym}
\usepackage{widetext}
\usepackage{cancel}
\usepackage[dvipsnames]{xcolor}
\newcommand{\del}[2]{\frac{\partial}{\partial #1}#2}
\title[Stability of Rotating Self-Gravitating Filaments]{Stability of Rotating Self-Gravitating Filaments: Effects of Magnetic Field}
\author[ Shubhadeep Sadhukhan, Surajit Mondal and Sagar Chakraborty]{Shubhadeep Sadhukhan$^{1}$\thanks{E-mail: deep@iitk.ac.in}, Surajit Mondal$^{2}$ \thanks{E-mail:
surajit@ncra.tifr.res.in}, and Sagar
Chakraborty$^{1,3}$\\
$^{1}$Department of Physics, Indian Institute of Technology Kanpur, U.P.-208016, India.\\
$^{2}$National Centre for Radio Astrophysics, TIFR, Pune, Maharashtra, 411007, India.\\
$^{3}$Mechanics \& Applied Mathematics Group, Indian Institute of Technology Kanpur, U.P.-208016, India.}
\begin{document}
\pagerange{\pageref{firstpage}--\pageref{lastpage}} \pubyear{2002}

\maketitle

\label{firstpage}

\begin{abstract}
We have performed systemmatic local linear stability analysis on a radially stratified infinite self-gravitating cylinder of rotating plasma under the influence of magnetic field. {In order to render the system analytically tractable, we have focussed solely on the axisymmetric modes of perturbations.} Using cylindrical coordinate system, we have derived the critical linear mass density of a non-rotating filament required for gravitational collapse to ensue in the presence of azimuthal magnetic field. Moreover, for such filaments threaded by axial magnetic field, we show that the growth rates of the modes having non-zero radial wavenumber are reduced more strongly by the magnetic field than that of the modes having zero radial wavenumber.  More importantly, our study contributes to the understanding of the stability property of rotating astrophysical filaments that are more often than not influenced by magnetic fields. In addition to complementing many relevant numerical studies reported the literature, our results on filaments under the influence of magnetic field generalize some of the very recent analytical works (e.g.,~\citet{jog2014}, etc.). For example, here we prove that even a weak magnetic field can play a dominant role in determining stability of the filament when the rotation timescale is larger than the free fall timescale. A filamentary structure with faster rotation is, however, comparatively more stable for the same magnetic field. {The results reported herein, due to strong locality assumption, are strictly valid for the modes for which one can ignore the radial variations in the density and the magnetic field profiles. %Arguably, the destabilizing role of the magnetic tension in the azimuthal magnetic field should be more prominent in a global analysis \`a la~\citet{Fiege01012000} which would be a natural extension of our paper.}  
}
\end{abstract}
\begin{keywords}
gravitation -- magnetohydrodynamics -- instabilities -- cosmology: large-scale structure of the Universe -- ISM: structure -- galaxies: clusters: intracluster medium
\end{keywords}
\section{Introduction}
Large scale filamentary structures are quite commonly found in our universe.
They are present in interstellar media \citep{andre2010, arzoumanian2011, jackson2010, kirk2013, palmeirim2013}.
These structures also connect galaxy clusters \citep{bond1996, springel2005, Fuma2011} and form a gas reservoir for the galaxies to accrete and grow \citep{keres2005, dekel2009}.
Numerical simulations suggest that galaxy clusters and cloud cores are formed at the intersection points of these filaments \citep{gray2013}.
\citet{andre2010,andre2014} have proposed that turbulence-driven filaments in the interstellar medium may be first stages of star formation.
Simulations also suggest that filamentary structures arise either due to turbulence in astrophysical fluids \citep{padoan2001}
or during intermediate stages of gravitational collapse \citep{gomez2014}.
Given the prevalent interest in the formation and the dynamics of the filamentary structures, it is an important exercise to explore the stability properties of the filaments in their realistic settings.
Both simulations and observations suggest that the inner cores of all filaments are composed of gas. 
Hence we can treat them as self-gravitating cylinders in hydrostatic equilibrium.
The stability of the cylindrical structures are not so well studied as the spherical ones. 
\citet{chandrasekhar1953} have studied the stability of an infinite incompressible self-gravitating cylinder with a constant axial magnetic field.
\citet{ostriker1964b} has extended their work for a compressible cylinder.
\citet{simon1963} and \citet{stodolkiewicz1963} have deduced an expression for the minimum length-scale (critical length) required for instability to set in.
\citet{mikhailovskii1972}, \citet{mikhailovskii1973}, and \citet{fridman1984} have studied the instabilities in a self gravitating cylinder of finite radius and infinite length.
But they have assumed a homogeneous density profile which does not match with an isothermal self-gravitating cylinder \citep{ostriker1964}.
\citet{nagasawa1987} has derived the dispersion relation for an infinite self-gravitating isothermal cylinder with an axial magnetic field and has found a critical length.
Later~\citet{Fiege01012000} have numerically studied the global stability of non-rotating self-gravitating isothermal filaments threaded with helical magnetic fields having radial dependence.
More complex situations like fragmentation of infinite filaments~\citep{bastien1983,arcoragi1991, bastien1991, nakamura1993, matsumoto1994, tomisaka1995, nakamura1995, tomisaka1996},  different equations of states~\citep{bastien1979}, effect of magnetic field and turbulence on framentation~\citep{seifried2015}  were also studied numerically.
\citet{tomisaka1995} and \citet{hanawa2015} have studied the stability of self-gravitating isothermal filament in presence of a magnetic field perpendicular to the axis of the filament.
The critical mass, which is defined as the minimum mass which is required to initiate a runaway collapse of the filament, was obtained numerically.
However, observations suggest that most of the filaments are far from being isothermal \citep{stepnik2003,palmeirim2013}.
Hence in order to study the stability of a realistic filament, it would be more convenient and useful to have a general dispersion relation that holds for a rather general class of density profiles and equations of state.
\citet{schneider2011} and \citet{breysse2014} 
have done detailed analytical calculations investigating the stability for non-rotating filaments assuming a more general polytropic equation of state.
For a rotating infinite cylinder having an arbitrary density profile and a polytropic equation of state, \citet{jog2014}
 have investigated its local and linear stability.
It must be emphasized that in the aforementioned works, the effects of magnetic field have not been analytically studied even though such effects are known to have crucial impact on the stability of the filaments \citep{tilley2007,Li2010,hennebelle2013}. 
Therefore, it is worthwhile to find out these effects analytically and among other things, to validate some of the existing numerical results in this context while generalising them with a view to investigating the interesting interplay between the rotation and the magnetic field.
To this end, this paper systemmatically explores the combined effect of the magnetic field and the rotation on the stability of the filamentary structures.
In the literature there is a relative dearth of papers which deals with the simultaneous influence of the rotations and the magnetic field on the dynamical stability of a self-gravitating filament, and thus, the present paper is a very important step in this direction.
\section{Linear stability analysis}

\begin{figure}
\centering
\scalebox{.35}{\includegraphics{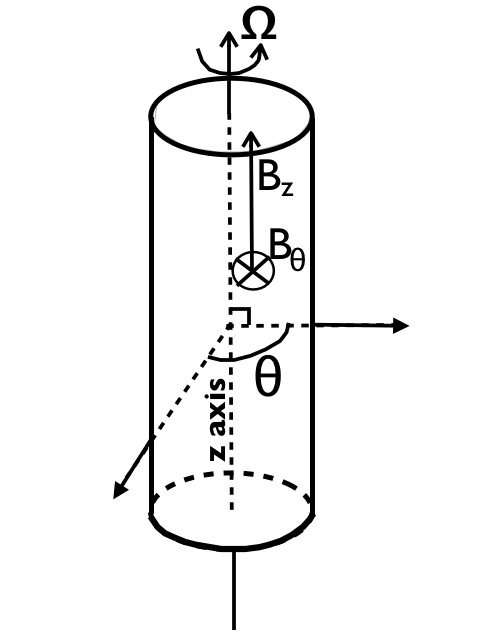}}  
\caption{\small \textit{Schematic diagram:} The gas cylinder is infinite in extent in $\pm z-$directions. At any point in the filament the unperturbed magnetic field has two components --- one along $z$-axis and the other one along azimuthal direction (into the plane of the paper for the point shown in the figure). The filament is differentially rotating about $z-$axis.}
\label{f:schematic}
\end{figure}

As depicted in Fig.~\ref{f:schematic} we model the filament as an infinite, self-gravitating, rotating cylinder.
We wish to study its stability in the presence of a uniform axial and azimuthal magnetic field.
We assume that the pressure and density in the cylinder are related by a polytropic equation of state.
A cylindrical geometry favours the growth of local perturbations before global modes take over \citep{pon2012}.
Hence we have assumed an unperturbed system in equilibrium and neglected the global collapse of filamentary cloud.
Although tides are known to affect astrophysical fluid instabilities \citep{jog2014, mondal2015}, for simplicity we neglect any such external agent.

\subsection{Unperturbed system}

\noindent We assume that the unperturbed system has an azimuthal symmetry. 
In equilibrium state density, pressure, and angular velocity depend only on radial distance.
The equilibrium density is denoted by $\rho_0$, the pressure by $P_0$ and the gravitational potential by $\phi_0$.
We have assumed that all particles has an angular velocity~($\Omega_0(r) \mathbf{\hat{e}_z}$) directed parallel to the symmetry axis. So the unperturbed velocity is $\textbf{v}_{0}=r\Omega_{0}(r) \mathbf{\hat{e}_\theta}$.
At equilibrium the cylinder is threaded by a uniform magnetic field $\mathbf{B_0}=B_\theta \mathbf{\hat{e}_\theta} + B_z \mathbf{\hat{e}_z}$.
z axis is directed along the axis of the cylinder.
We treat the system as an inviscid magnetofluid and assume that the electrical resistivity is negligible.
The equations governing the dynamics of the system are given by,
\begin{align}
& \label{e:contd} \frac{\partial \rho}{\partial t} + \nabla \cdot (\rho \mathbf{v})  =0\, ,\\
& \frac{\partial \mathbf{v}}{\partial t}+ \mathbf{(v \cdot \nabla)v}  =-{\nabla h} - \nabla \phi -\frac{\mathbf{B \times (\nabla \times B)}}{4 \pi {\rho}} \label{e:motion1}\,,\\
& \frac{\partial \mathbf{B}}{\partial t}= \nabla \times (\mathbf{v} \times \mathbf{B}) \label{e:flux_freeze}\,,\\
& \label{e:poisson1} \nabla^2 \phi = 4 \pi G \rho\, .
\end{align}
{ h is the enthalpy and is defined as $dh=\rho^{-1} dp$}. It should be noted that equations (\ref{e:contd}) to (\ref{e:poisson1}) are written in an inertial frame.
The perturbed density, pressure, velocity, magnetic field and gravitational potential are given by:
$\rho  =\rho_0+\rho_1$,
{$h=h_0+h_1$},
$\mathbf{v}  = \mathbf{v_0+v_1}$,
$\mathbf{B}  =\mathbf{B_0 +B_1}$, and 
$\phi  =\phi_0+\phi_1\,.$
Here $\rho_1$, ${h_1}, \mathbf{v_1}, \mathbf{B_1},$ and $ \phi_1 $ are infinitesimal perturbations; and $\rho_0$, ${h_0}, \mathbf{v_0}, \mathbf{B_0},$ and $ \phi_0 $ are the equilibrium solutions.
$\mathbf{v}_0$ and $\mathbf{B}_0$ have already been mentioned earlier; and $\rho_0$, ${h_0},~and~$ $\phi_0 $ can be determined from the following equations:

\begin{align}
 r \Omega_0^2= & ~{\frac{\partial h_0}{\partial r}}+\frac{\partial \phi_0}{\partial r}\label{e:force}\,,\\
\nabla^2 \phi_0= & 4 \pi G \rho_0\label{e:poisson2}\,.
\end{align}

Pressure and density are related by a simple power law, $P\propto \rho^m$ (polytropic process).
This relation can closely characterize a wide range of thermodynamic processes. $m$ can range from $0$ to $\infty$ which includes isobaric ($m=0$), isothermal ($m=1$), isoentropic ($m=\gamma$), and isochoric ($m=\infty$) processes.
$\gamma=5/3$ for ideal monoatomic gas.

\subsection{Dispersion Relation}
Linearising equations (\ref{e:contd})-(\ref{e:poisson1}) about the equilibrium state we get,
\begin{align}
& \frac{\partial \rho_1}{\partial t}+ \nabla \cdot \rho_1 \mathbf{v_0}+\nabla \cdot \rho_0 \mathbf{v_1}=0 \label{e:continuity},\\
 \label{e:motion} & \frac{\partial \mathbf{v_1}}{\partial t} =\begin{aligned}[t] - & \mathbf{(v_0 \cdot \nabla)v_1} - \mathbf{(v_1 \cdot \nabla)v_0} -{\nabla h_1} -  \nabla \phi_1 \\  & -\frac{\mathbf{B_0 \times (\nabla \times B_1)}}{4 \pi \rho_{0}}-\frac{\mathbf{B_1 \times (\nabla \times B_0)}}{4 \pi \rho_{0}}, \end{aligned}\\
& \frac{\partial \mathbf{B_1}}{\partial t}= \nabla \times (\mathbf{v_0} \times \mathbf{B_1}) + \nabla \times (\mathbf{v_1} \times \mathbf{B_0}) \label{e:flux_freezing},\\
& \nabla^2 \phi_1 = 4 \pi G \rho_1 \label{e:poisson}.
\end{align}
For convenience we shall focus on the growth of axisymmetric perturbations i.e.,~the perturbations in the form
\begin{equation}
 \label{e:perturbation} f_1=\widehat{f_1} e^{i(k_r r +k_z z -\omega t)}. 
\end{equation}
Also, we assume that the perturbations are local~{\citep{toomre1964}}. 
Hence $k_r r_0\gg1$, where $r_0$ is the typical radius of the unperturbed distribution within which the perturbation occurs.
The physical implication of this assumption is that the length scale over which the unperturbed quantities vary is much larger than the length scale over which perturbed quantities vary spatially.
Under this assumption,
\begin{align}
& \del{r}{\left (\frac{P_1}{\rho_0} \right )} =\del{r}{\left(\frac{c_0^2 \rho_1}{\rho_0}\right )} \approx \frac{c_0^2}{\rho_0} \frac{\partial \rho_1}{\partial r}\,, \\
& \frac{1}{r} \del{r}{(r \rho_0 v_{1r})} \approx \rho_0 \frac{\partial v_{1r}}{\partial r}\,, \\
& \frac{1}{r} \del{r}{\left (r \frac{\partial \phi_1}{\partial r}\right )} =\frac{\partial^2 \phi_1}{\partial r^2}+\frac{1}{r} \frac{\partial \phi_1}{\partial r} \approx \frac{\partial^2 \phi_1}{\partial r^2} \,,\\
& \frac{1}{r} \del{r}{(r B_{1\theta})}=\frac{B_{1\theta}}{r}+\frac{\partial B_{1\theta}}{\partial r} \approx \frac{\partial B_{1\theta}}{\partial r} \,.
\end{align}
\\
\\
\\
\noindent We substitute equation (\ref{e:perturbation}) in equations (\ref{e:continuity})-(\ref{e:poisson}) and using the aforementioned locality assumption we arrive at the following matrix equation:

\begin{widetext}	
\begin{eqnarray}  \label{e:eq}
\begin{pmatrix}
\frac{\rho_0 k_r}{\omega} & 0 & \frac{\rho_0 k_z}{\omega} & -1 & 0\\
0 & 0 & 0 & \frac{4 \pi G}{k^2} & 1\\
\omega+\frac{i}{4\pi \rho_0} ( \frac{B_\theta R_4}{r}-B_z P_4+B_\theta P_7 )& -2i\Omega_0+\frac{i}{4\pi \rho_0} (\frac{B_\theta R_5}{r}+B_\theta P_8 )& \frac{i}{4\pi \rho_0} (\frac{B_\theta R_6}{r}+B_\theta P_9) & -\frac{k_r c_0^2}{\rho_0}& -k_r\\

-2iY+\frac{i}{4\pi \rho_0} (B_z P_1-\frac{B_\theta R_1}{r})& \omega+\frac{iB_z P_2}{4\pi \rho_0} & \frac{iB_z P_3}{4\pi \rho_0} & 0& 0\\

-\frac{iB_\theta P_1}{4\pi\rho_0}& -\frac{iB_\theta P_2}{4\pi \rho_0}& \omega-\frac{iB_\theta P_3}{4\pi \rho_0}& -\frac{k_z c_0^2}{\rho_0}& -k_z
\end{pmatrix}
\begin{pmatrix}
\widehat{v}_{1r}\\
\widehat{v}_{1\theta}\\
\widehat{v}_{1z}\\
\widehat{\rho}_1\\
\widehat{\phi}_1
\end{pmatrix}
=0.
\end{eqnarray}
\end{widetext}
\noindent Here 
\begin{align*}
& R_1\equiv -\frac{k_z B_z}{\omega}, R_4\equiv \frac{2ik_z B_z(Y+\Omega_0)}{\omega^2}+\frac{k_r B_\theta}{\omega},\, R_5\equiv-\frac{k_z B_z}{\omega}\,,\\
& R_6\equiv\frac{k_z B_\theta}{\omega}\, R_7\equiv\frac{B_z k_r}{\omega r},\, P_1\equiv -iR_4 k_z,\, P_2\equiv \frac{ik_z^2 B_z}{\omega},\\
& P_3\equiv-\frac{ik_z^2 B_\theta}{\omega},\,P_4\equiv -\frac{iB_z k^2}{\omega},\, P_7\equiv -\frac{k_r P_1}{k_z},\, P_8\equiv -\frac{ik_r k_z B_z}{\omega},\,\\
&  P_9\equiv -\frac{B_\theta P_8}{B_z}\,,Y\equiv-\frac{1}{2} \left[ \Omega_0(r) +\del{r}{(r \Omega_0 ( r ) )}\right ]\,,\, k^2\equiv k_z^2+k_r^2\,.
\end{align*}
The effective sound velocity ($c_0$) is given by $c_0^2=(\partial P_0 /\partial \rho_0)_{T_0}$. 
For equation (\ref{e:eq}) to have non-trivial solution, it is required that
\\
\\
\\

\begin{widetext}
\begin{equation}
\begin{vmatrix}
\frac{\rho_0 k_r}{\omega} & 0 & \frac{\rho_0 k_z}{\omega} & -1 & 0\\
0 & 0 & 0 & \frac{4 \pi G}{k^2} & 1\\
\omega+\frac{i}{4\pi \rho_0} ( \frac{B_\theta R_4}{r}-B_z P_4+B_\theta P_7 )& -2i\Omega_0+\frac{i}{4\pi \rho_0} (\frac{B_\theta R_5}{r}+B_\theta P_8 )& \frac{i}{4\pi \rho_0} (\frac{B_\theta R_6}{r}+B_\theta P_9) & -\frac{k_r c_0^2}{\rho_0}& -k_r\\

-2iY+\frac{i}{4\pi \rho_0} (B_z P_1-\frac{B_\theta R_1}{r})& \omega+\frac{iB_z P_2}{4\pi \rho_0} & \frac{iB_z P_3}{4\pi \rho_0} & 0& 0\\

-\frac{iB_\theta P_1}{4\pi\rho_0}& -\frac{iB_\theta P_2}{4\pi \rho_0}& \omega-\frac{iB_\theta P_3}{4\pi \rho_0}& -\frac{k_z c_0^2}{\rho_0}& -k_z
\end{vmatrix}=0\,,\label{e:det}
\end{equation}
\end{widetext}
which is the dispersion relation for the system in hand.

\section{Results}
In this section, we use equation (\ref{e:det}) to study the effect of the magnetic fields on the stability of filaments in different situations. 
We also compare our results with the ones already reported in the literature wherever possible.
\subsection{Axial magnetic field, no rotation} \label{s:nagasawa}
As the first case we consider the simplest situation of a non-rotating filament threaded only be a constant axial magnetic field.
From equation (\ref{e:det}), in absence of rotation and $B_\theta$, the dispersion relation is given by,
\begin{eqnarray}
 \label{e:disp} \omega^6-\omega^4 \left[ \omega_j^2 + v_A^2 \left ( k^2+k_z^2 \right ) \right ]+  \omega^2 \left[ v_A^4 k_z^2 k^2+ 2 \omega_j^2 v_A^2 k_z^2 \right ]\nonumber\\ 
 -v_A^4 k_z^4 \omega_j^2 =0\,,
\end{eqnarray}
where $\omega_j^2\equiv c_0^2 k^2-4 \pi G \rho_0$ and $v_A^{2} \equiv {B_0^{2}}/{4\pi \rho_0}$.
$v_A$ is Alfv\'en velocity. $\omega_j$ is the growth rate of Jeans instability (in absence of rotation and magnetic field) wherein perturbations with wavelength larger than $\lambda_j \equiv c_0 \sqrt{{\pi}/{G\rho_0}}$ results in gravitational collapse.
Note that $\omega_j^2$ is negative for the system to be unstable in the absence of the rotation and the magnetic field.

The solution to this equation in the absence of the magnetic field is $\omega_j$.
Including the first order correction due to the presence of the magnetic field, {{as described in Appendix~A,}} yields,
\begin{equation}
\omega^2=\omega_j^2+{v_A^2 k_r^2}\,.\label{e:nagaxdisp}
\end{equation}
From this equation it is clear that for a small magnetic field, the growth rate of pure instability is decreased by the magnetic field.
This conclusion is qualitatively in line with \citet{nagasawa1987} but quantitatively differs from it because he worked with non-axisymmetric perturbations having $k_r=0$. 
More precisely, our result highlights the fact that the growth rates of the modes having non-zero radial wavenumber are decreased more strongly by the magnetic field than that of the modes having zero radial wavenumber.
Note that $k_z$ can appear in the perturbative expression of $\omega^2$ only at higher order corrections than what has been presented in equation (\ref{e:nagaxdisp}).    

In hindsight that the magnetic field should affect the stability criterion seems quite natural.
\citet{elmegreen1982}
has derived a criterion for stability in self-gravitating exponential gas layer in the presence of transverse magnetic.
In the filamentary system under consideration, the density stratification is along $r$ and the magnetic field is along $z$.
Due to locality assumption the qualitative features of the stability criterion are expected to be independent of the cylindrical (global) geometry.
Hence qualitatively the stability criterion of our system is same as that obtained in \citet{elmegreen1982} in the sense that in both cases magnetic field affect the stability criteria.

In passing, we mention that we have done the calculation for the case of non-axisymmetric perturbations as well and it is easy to show that the dispersion relation remains similar in form to equation (\ref{e:disp}). The only change is that in this case $k^2=k_r^2+k_\theta^2/r^2+k_z^2$. Therefore, the corresponding dispersion relation has an extra first order correction viz., $v_A^2 k_\theta^2/r^2$ added to the R.H.S. of equation (\ref{e:nagaxdisp}). 
\subsection{Azimuthal magnetic field, no rotation}
Equation (\ref{e:det}), in the absence of the rotation and $B_z$, gives the following dispersion relation:

\begin{equation} \label{e:disp_perp_mag}
\omega^2={{\omega_j^2+v_A^2k^2}}\,.
\end{equation}
{ Equation~(\ref{e:nagaxdisp}) and equation~(\ref{e:disp_perp_mag}) resemble each other quite closely; the only difference is that in the former case $k_r^2$ shows up whereas $k^2$ appears in the latter case. This is not coincidental at all. Had we found dispersion relation~(\ref{e:disp_perp_mag}) using $\mathbf{k}=k_z\hat{\mathbf{e}}_z$ and also rewrite equation~(\ref{e:nagaxdisp}) by putting $\mathbf{k}=\frac{k_\theta}{r}\hat{\mathbf{e}}_\theta$, then we would have obtained exactly identical-looking dispersion relations, viz., $\omega^2=\omega_j^2+v_A^2k^2$. Note that in both these cases, the three physical vector quantities magnetic field, acceleration due to gravity, and wave vector are mutually perpendicular, thus, locally creating equivalent configurations.}

We substitute for $v_A$ and then solve for the critical density $\rho_{0_{crit}}$~(i.e.,~when $\omega=0$) to get,
\begin{equation}
\rho_{0_{crit}}=\frac{c_0^2 k^2 + Gk^2 \sqrt{\frac{c_0^4}{G^2}+\frac{4B^2}{Gk^2}}}{8 \pi G}\label{e:rho_crit}.
\end{equation}
Here we have ignored the unphysical root for $\rho_{0_{crit}}$.
We substitute $\rho_{0_{crit}}=M_{crit}/Al$, where $A$ is the cross-section area of cylinder and $l$ is its length.
$l$ can be chosen to be the length scale over which the equilibrium quantities change.
We also take $A\sim l^2$ and $k\sim 2\pi/l$ for our order of magnitude estimate.
Then we arrive at,

\begin{equation}
\lambda_{crit}\equiv\frac{M_{crit}}{l}=\pi \Bigg [ \frac{c_0^2}{2G}+\sqrt{\bigg(\frac{c_0^2}{2G}\bigg)^2+\bigg (\frac{\Phi}{2\pi \sqrt{G}} \bigg)^2} \Bigg ],
\end{equation}

\noindent where $\Phi$ is the magnetic flux per unit length threading through the box and $\lambda_{crit}$ is the critical linear mass density.
Apart from the order unity numerical coefficient --- specifically speaking, $\pi$ outside the square brackets --- the critical mass density obtained by \citet{hanawa2015} matches with our result.
{{It is easy to check that the mismatch of $\pi$ stems from the fact that their analysis, which uses Cartesian coordinates against our choice of cylindrical coordinates, considers a two-dimensional `flattened filamentary cloud' to be confined in $z=0$ plane and to have Lorentz profile for density along $x$-axis. We, however, are working with a three-dimensional cylindrical filament.}}
\subsection{Axial magnetic field, rotation included}
\begin{figure*}[h!]
	\scalebox{0.55}{\includegraphics[trim=1cm 0cm 0cm 0cm]{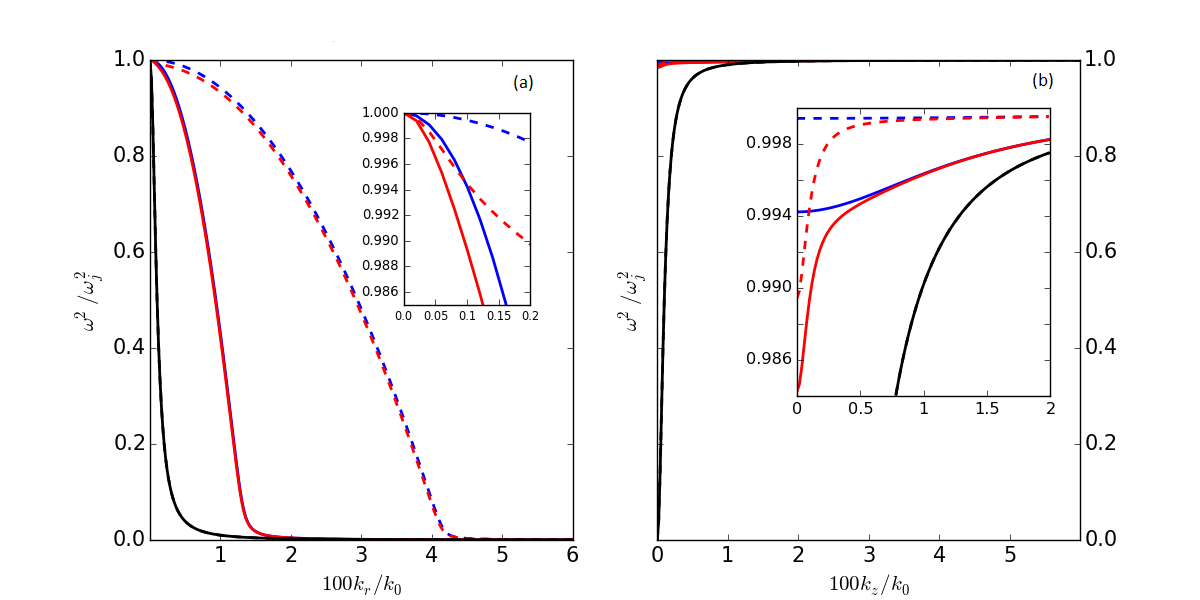}}
	\caption{(\textit{Colour online}) (a) Growth rate squared $\omega^2$ vs radial wavenumber $k_r$, and (b) $\omega^2$ vs axial wave number $k_z$ for different values of $\kappa$--- $0$~s$^{-1}$~(blue curves), $10^{-16}$~s$^{-1}$~(red curves) and $10^{-14}$~s$^{-1}$ \citep{recchi2014}~(black curves); and for two different values of $v_A^2$--- $10^2$~$m^2s^{-2}$~(dashed lines) and $10^3$~ $m^2s^{-2}$~(solid lines) \citep{ferland2011}. $\omega_j^2$ has been kept constant at $-10^{-30}\: \mathrm{s}^{-2}$ \citep{Zuckerman19742, Zuckerman19741, pon2012} and we have chosen $k_0=74\:\mathrm{pc}^{-1}$ \citep{jog2014}. In left panel~(a), $k_z=0.001\:k_0$, whereas for right panel~(b), $k_r=0.001\:k_0$. In both the panels, the black dashed curves are not visible as they coincide with the respective black solid curves.}
	\label{f:ovsk}
\end{figure*}
When both the axial magnetic field and the rotation are simultaneously turned on, the situation is way more complex.
In this case, the dispersion relation is given by,
\begin{equation} \label{e:disp_rot}
\begin{split}
& \omega^6-\omega^4 \left[ \left (\kappa^2+\omega_j^2 \right ) + v_A^2 \left ( k^2+k_z^2 \right ) \right ] + \omega^2 \biggl[ \frac{\kappa^2 k_z^2 \omega_j^2}{k^2}\\&+v_A^4 k_z^2 k^2+v_A^2 k_z^2 (\kappa^2+2\omega_j^2-{4\Omega_0^2} )\biggr] -v_A^4 k_z^4 \omega_j^2\\
&-\frac{v_A^2k_z^4\omega_j^2(\kappa^2-{4\Omega_0^2})}{k^2}=0,
\end{split}
\end{equation}
where $\kappa^2\equiv-4 \Omega_0 Y$.
We have plotted a few representative numerical solutions of this dispersion relation in Fig.~\ref{f:ovsk}. { In our system, rotation can act in addition to the magnetic pressure to stabilize the filament as expected. It is, thus, not too surprising to see from the plots, that while $k_z\rightarrow\infty$ limit recovers a purely Jean infinite medium with only pressure support against gravity, for $k_r\rightarrow\infty$ limit the fluid relaxes into rotational support everywhere yielding a vanishing growth rate.} Careful study of these plots bring forward the rich physics present due to the interplay of three different timescales --- free-fall timescale ($1/|\omega_j|$), rotation timescale ($1/\kappa$) and Alfv\'en timescale $1/v_Ak$ --- relevant in the problem of gravitational collapse of the filaments. Being interested in perturbative effect of the weak magnetic field present near astrophysical filaments, we keep the Alfv\'en timescale the largest. We systematically inspect the plots for two different cases: $\kappa^2<-\omega_j^2$ and $\kappa^2>-\omega_j^2$, and observe the following facts:

\begin{enumerate}
\item For both the cases, increasing $k_z$ (fixed $k_r$) causes the growth rate of instability to monotonically increase to a value close to $\omega_j$.
\item Again, for both the cases, increasing $k_r$ (fixed $k_z$) monotonically decreases the growth rate of the instability to a vanishingly small value.
\item A fixed absolute increase in axial magnetic field is more effective in stabilising the gravitational collapse (i.e.,~reducing the growth rate) for $\kappa^2<-\omega_j^2$ than for the case $\kappa^2>-\omega_j^2$.

\end{enumerate} 

These features of the curves can be analytically justified by taking the effect of the magnetic field perturbatively.
The solution to equation (\ref{e:disp_rot}) in the absence of the magnetic field is given by $\omega_0$ and this solution has been provided in \citet{jog2014}.
Including the first order correction due to the presence of the magnetic field, the solution is given by,
\begin{equation}
\omega^2=\omega_0^2+\omega_1^2\left[\frac{\kappa^2k_z^2\omega_j^2}{k^2}-2(\kappa^2+\omega_j^2)\omega_0^2+3\omega_0^4\right]^{-1}\,,\label{e:rot_lin}
\end{equation}
where
\begin{align}
& \omega_0^2\equiv \frac{1}{2}\left[b\pm\sqrt{b^2-4c}\,\right]\,,\,\, \left(b\equiv\omega_j^2+\kappa^2\,,\,c\equiv \frac{\kappa^2 k_z^2 \omega_j^2}{k^2}\right)\,;\nonumber\\
&\omega_1^2\equiv\begin{aligned}[t]&\omega_0^4\left[v_A^2(k^2+k_z^2)\right]-\nonumber\\&\biggl[v_A^2 k_z^2(\kappa^2  + 2\omega_j^2-{4\Omega_0^2})\biggr]\omega_0^2+\frac{v_A^2k_z^4\omega_j^2}{k^2}(\kappa^2-{4\Omega_0^2})\,.\end{aligned}\nonumber
\end{align}
Now we consider following two different cases in which only the negative value of $\omega_0^2$ will be considered because our motivation in this paper is to see how much the magnetic field stabilises an unstable scenario.
\subsubsection{$-\omega_j^2\gg\kappa^2\gg v_A^2 k^2$}
\noindent Noting that
\begin{align}
& \omega_0^2-\omega_j^2\approx \kappa^2k_r^2/k^2\,,\\
&\omega_1^2 \approx \omega_j^4 v_A^2 k_r^2+\frac{\omega_j^2 \kappa^2 v_A^2}{k^2}\left[2k_r^4+k_r^2 k_z^2\left(1+{\frac{4\Omega_0^2}{\kappa^2}}\right)\right]\,,\textrm{ and}\\
& 3\omega_0^4-2\omega_0^2(\omega_j^2+\kappa^2)+\frac{\kappa^2 \omega_j^2 k_z^2}{k^2}\approx  \omega_j^4+\frac{\omega_j^2 \kappa^2}{k^2} (2k_r^2-k_z^2)\,,
\end{align}
upto ${\kappa^2}/{\omega_j^2}$ corrections, equation (\ref{e:rot_lin}) can be simplified to arrive at
\begin{align}
& \omega^2 \approx \omega_j^2+\frac{\kappa^2}{1+k_z^2/k_r^2}+v_A^2k_r^2\left[1+\frac{2k_z^2\kappa^2}{k^2\omega_j^2}\left(1+{\frac{2 \Omega_0^2}{\kappa^2}}\right) \right]\,.\label{e:small_kappa}
\end{align}
From this expression it is clear that the instability rate will decrease on increasing $k_r$ ($k_z$ being fixed) as is seen in Fig.~\ref{f:ovsk}(a). Also we see that monotonic nature of the growth rate is validated by equation (\ref{e:small_kappa}). In case we fix $k_r$ and vary $k_z$, we observe from Fig.~\ref{f:ovsk}(b) that the relevant curves start from a positive value, increases slowly and asymptotically reaches a value close to unity. From equation (\ref{e:small_kappa}) it is clear that this value is $(\omega_j^2+v_A^2k_r^2)/\omega_j^2$ (upto the first subleading correction).
\subsubsection{$\kappa^2\gg-\omega_j^2\gg v_A^2 k^2$}
It is known that the angular frequency of star-forming filamentary structures can be as high as $6.5\cdot10^{-14}\,\textrm{s}^{-1}$ \citep{recchi2014}. 
Though possibly this case of very high rotation rate is less probable, we should not exclude it from our discussion as it helps us to understand the filamentary system in great details.
Moreover, $\kappa$ contains a term containing gradient of $\Omega_0(r)$ that can make $\kappa$ larger. 
Anyway, the expression for $\omega_0^2$, in the present ordering of timescales, reduces to:
\begin{equation}
\omega_0^2 \approx \frac{\omega_j^2}{1+k_r^2/k_z^2}\,.
\end{equation}
Also,
\begin{equation}
\frac{\kappa^2 k_z^2 \omega_j^2}{k^2}-2(\kappa^2+\omega_j^2)\omega_0^2+3\omega_0^4 \approx -\frac{\omega_j^2\kappa^2 k_z^2}{k^2}\,.
\end{equation}
Substituting these expressions in equation (\ref{e:rot_lin}), we get,
\begin{equation}
\omega^2\approx\frac{\omega_j^2}{1+k_r^2/k_z^2}\left[1+\frac{v_A^2k_r^2}{\kappa^2}\right]\label{e:high_kappa}\,.
\end{equation}
For fixed $k_r$, we note that the growth rate is mainly determined by $(1+k_r^2/k_z^2)^{-1}$-term as $\kappa$ is very large.
Therefore, on increasing $k_z$ the instability's growth rate increases as demonstrated in Fig.~\ref{f:ovsk}(b) and approaches a value close to $|\omega_j|$.
Also it is obvious from equation~\ref{e:high_kappa} that an increase in $k_r$ (for fixed $k_z$) decreases $\omega^2$. 
Moreover, it is clear by comparing equations (\ref{e:small_kappa}) and (\ref{e:high_kappa}) that $\omega$ in the present case has relatively weaker dependence on the magnetic field and indeed that is what we observe in Fig.~\ref{f:ovsk}(a)-(b) --- the black solid and the dashed lines coincide.
This is because $v_A^2k_r^2$ appears along with $\omega_j^2/\kappa^2$ as a multiplicative factor unlike in equation (\ref{e:small_kappa}) where it appears with unity as the multiplicative factor.
Hence, unless the magnetic field changes by a relatively large amount, one will not observe any appreciable change in the instability rate.
In other words, a fixed change in axial magnetic field is more effective in decreasing the growth rate of the instability for the case where $\kappa^2<-\omega_j^2$ than for the case where $\kappa^2>-\omega_j^2$. 
%
%%%%%%%%%%%%%%%%%%%%%
\section{Discussions and Conclusions}
In this work we have derived an analytical expression for the dispersion relation describing the linear stability of a self-gravitating rotating infinite cylindrical filament under the influence of magnetic field. We have shown mathematically how the magnetic field, in general, reduces the growth rate of instability that leads to gravitational collapse. This has also been clearly illustrated and extensively discussed using Fig.~\ref{f:ovsk}.

Specifically, we have seen that magnetic field plays a dominant role in stabilising the filament, more so when the rotation timescale is larger than the free fall timescale.
Also, in the absence of the rotation, we have found out an explicit expression for the linear critical mass density as a function of the sound speed and magnetic flux. 
Moreover, our results show what by taking $k_r=0$, \citet{nagasawa1987} overlooked the modes whose growth are more strongly obstructed by the magnetic field. We have shown that the growth rates of the modes having non-zero radial wavenumber are decreased more strongly by the axial magnetic field than that of the modes having zero radial wavenumber.     
Actually Fig.~\ref{f:ovsk} and the related detailed calculations lend us following intuitive way of looking at the filament's stability: Owing to centrifugal force and Lorentz force respectively, both the axial rotation and the axial magnetic field are expected to counteract gravitational collapse, and hence the rate of growth of instability is reduced. There is, however, an interesting aspect to it--- the rotation rate always pushes the fluid outwards whereas the magnetic field presents restoring force to both the radially outgoing and the radially incoming matter. Hence, it may be expected that the magnetic field can in principle counteract the stabilising effect of the rotation by halting the rotation-induced radial outflow of the matter.

{It should, however, be noted that the locality assumption, i.e., $\mathbf{k}\cdot \mathbf{r} \gg 1$, --- although very convenient for doing linear stability analysis analytically for the system in hand--- is a limitation of our analysis. Under this assumption we have managed to ignore the realistic radial dependence of the equilibrium profiles of the density and the magnetic field. Needless to say, one expects both of these physical quantities to become weaker with the distance from the axis of the filament, an effect that is amenable to global stability analysis~\citep{Fiege01012000}. The magnetic field is seen to provide pressure support against gravitational pull. It is in the global analysis, where $\mathbf{k}\cdot \mathbf{r} \sim 1$, the role of magnetic tension owing to the azimuthal component of the magnetic field is expected to become even more prominent and destabilizing. The curvature of the magnetic field, due to a non-zero azimuthal component, provides a radial tension force leading to the subtle interplay between the angular velocity and the magnetic field in stabilising or destabilizing the filament's collapse. Nevertheless, in the local regions where the linear stability analysis has been done in this paper, the magnetic tension force is generated due to curvature introduced in the field lines by the perturbations and these play an important role in the way perturbations behave temporally.} 

One can easily foresee several possible extensions of the present work. We note that in order to keep the algebra tractable, we have focussed mainly on the stability of axisymmetric perturbations.
Of course, the case of non-axisymmetric perturbations is also of interest.
\citet{nagasawa1987} has reported a numerical work using non-axisymmetric perturbations.
Thus, as a future direction, one can build on the work presented herein with a view to encompassing the effects of the non-axisymmetric perturbations extensively.
{This particular direction potentially involves even richer physics than discussed in this paper if one combines it with global analysis and, subsequently, one can also study the effect of simultaneous presence of the rotation and the magnetic field on the process of filament fragmentation~\citep{nakamura1993}.}
Additionally, a more detailed analysis with both the azimuthal and the radial components of the magnetic field simultaneously included can also be pursued as a sequel of this work.
We have also assumed that the particles in the filament have no axial velocity. But even if there exists a non-zero constant mean axial velocity, we can always go in a suitable inertial reference frame where there is no axial velocity and our calculations hold. However, as is evident, a non-uniform mean axial velocity complicates the calculation and could be another direction that merits further investigation. In fact the velocity distribution can be quite complex. The particles can be even be accelerated \citep{peretto2014, zernickel2013} 
due to large scale longitudinal collapse. \citet{schneider2011} have shown numerically that longitudinal expansion of tidal tails has a stabilising effect. This issue also needs to be revisited and studied to estimate the simultaneous effect of the magnetic field and the rotation.
An even more challenging direction of research could be to pursue weakly non-linear stability analysis of the instability resulting in the gravitational collapse of the filament in the presence of both the rotation and the magnetic field.

\section*{Acknowledgements}

\noindent The authors thank Prateek Sharma for useful discussions. SC gratefully acknowledges financial support received through the INSPIRE faculty fellowship (DST/INSPIRE/04/2013/000365) awarded by Department of Science and Technology (Government of India). 
\appendix
{
\section{Illustration of the perturbative scheme}\label{App:AppendixA}
We are considering the effect of very weak magnetic field on the stability of filaments of rotating magnetofluid. This set-up allows us to derive the expression for the growth rate of instability perturbatively. In this appendix, we illustrate the perturbation method by deriving equation~(19) from equation~(18). 

Under weak magnetic field assumption, $v_A^2k^2 \ll \omega_j^2 $, it is natural to consider the dimensionless quantity $\epsilon\equiv v_A^2k^2/\omega_j^2$ as the perturbation parameter. We begin with the ansatz:
$\omega^2=\omega_0^2+\epsilon\omega_1^2+\epsilon^2\omega_2^2+\cdots$. The standard idea now is to calculate $\omega^2_is$ iteratively. Thus, on using the ansatz in the dispersion relation~(18) and keeping terms only upto the first order in $\epsilon$, we arrive at:
\begin{align}
	\begin{split}
		(\omega_0^6+3\epsilon\omega_0^4\omega_1^2)-\omega_0^4\left[\omega_j^2+\epsilon\omega_j^2\left(1+\frac{k_z^2}{k^2}\right)\right]-2\epsilon\omega_0^2\omega_1^2\omega_j^2\\
+\omega_0^2\left[2\epsilon\omega_j^4\frac{k_z^2}{k^2}\right]&=0\,.
	\end{split}
\end{align}
Equating the coefficient of $\epsilon^0$ in the L.H.S. of the above equation with its R.H.S., yields
\begin{equation}
%	\begin{split}
		\omega_0^6-\omega_0^4\omega_j^2=0\,
		\implies\omega_0^2=\omega_j^2\,.\label{aa}
%	\end{split}
\end{equation}
Similarly, consideration of the coefficient of $\epsilon^1$ gets us, after simplification, the following:
\begin{equation}
	\begin{split}
		\omega_1^2=\frac{\omega_j^2k_r^2}{k^2}\,,
	\end{split}
\end{equation}
where have used equation~(\ref{aa}). Therefore, the perturbative solution upto the first order in $\epsilon$ is
\begin{equation}
	\omega^2=\omega^2_0+\epsilon\omega^2_1=\omega_j^2+v_A^2k_r^2\,.
\end{equation}
This is nothing but equation~(19).
}
%
%%%%%%%%%%%%%%%%%%%%%%%%%%%%%
\bibliographystyle{mn2e}

\bibliography{Sadhukhan_etal_references}
\bsp

\label{lastpage}

\end{document}